\documentclass[reprint, amsmath, amssymb, numerical, apl]{revtex4-1}

\usepackage{lineno,hyperref}
\modulolinenumbers[5]

\usepackage[a4paper, margin=1in]{geometry}
\usepackage{graphicx}
\usepackage{textcomp}
\usepackage{amsmath}
\usepackage{booktabs}
\usepackage[scriptsize]{subfigure}
\usepackage{epstopdf}
\usepackage{setspace}

\let\Gamma\varGamma

\usepackage[svgnames]{xcolor}
\usepackage{hyperref}
\hypersetup{
    pdftitle={Influence of space charge on domain patterns and susceptibility in a rhombohedral ferroelectric film},
    pdfauthor={Wei Li Cheah, Nathaniel Ng and Rajeev Ahluwalia},
    pdfkeywords={space charge} {ferroelectric} {faceting} {phase field} {BiFeO3},
    linkcolor=red,
    urlcolor=blue,
    citecolor=brown,
    colorlinks=true
}

\usepackage{fancyhdr}
\newcommand{\sub}[1]{\ensuremath{_{\text{#1}}}}
\newcommand{\bfo}{{BiFeO}\sub{3}}

\begin{document}

\title{Influence of space charge on domain patterns and susceptibility in a rhombohedral ferroelectric film}

\author{Wei Li Cheah}
\email{cheahwl@ihpc.a-star.edu.sg}

\author{Nathaniel Ng}

\author{Rajeev Ahluwalia}

\affiliation{Institute of High Performance Computing, 1 Fusionopolis Way, \#16-16 Connexis North, Singapore 138632, Singapore}

\begin{abstract}
The presence of a space charge region induces an internal electric field within the charged region that, in a ferroelectric material, would rotate the polarisations to align with the field. The strength of the induced field would therefore determine the domain patterns and polarisation switching properties of the material. Using a phase-field model, we investigate the effect of charged layers in fully and partially depleted \bfo\ thin films in the rhombohedral phase. While the domain pattern in a charge-free \bfo\ film consists of only two polarisation variants, we observe complex patterns with four coexisting variants that form within the charged layers at sufficiently high induced fields. These variants form a head-to-head configuration with an interface that is either wavy or planar depending on the internal field strength, which is determined by the charge density as well as the thickness of the charged layer. For depletion layers with sufficient thickness, there exists a range of charge density values for which the interface is wavy, while at high densities the interface becomes planar. We find that films with wavy interfaces exhibit enhanced susceptibilities with reduced hystereses compared to the charge-free film. The results of our work suggest that introducing space charge regions by careful selection of dopant density and electrode materials can engineer domain patterns that yield a higher response with a smaller hysteresis.

\end{abstract}


\maketitle

\section{Introduction}

When a metal electrode is placed in contact with a semiconductor material, there may be a flow of charge carriers from the semiconductor to the metal, depending on the work function of the metal \cite{Damjanovic1998}. An accumulation of charge carriers results at the electrode/semiconductor interface, inducing an electrostatic potential that depletes these carriers in adjacent regions \cite{Zubko2006,Yang2012thesis}. This space charge region, or depletion layer, is typically formed near interfaces such as grain boundaries and interfaces between heterophases, as well as near surfaces. In ferroelectric materials which are regarded as wide band gap semiconductors, the presence of such regions may affect the properties of the material, in many cases detrimental for applications such as non-volatile memory devices and capacitors. It lowers the ferroelectric phase transition temperature and suppresses ferroelectricity in the space charge regions \cite{Bratkovsky2000}. Furthermore, the domain pattern that forms under the influence of space charge can substantially differ from those without \cite{Xiao2005,Ng2012,Ahluwalia2013,Wang2012}. Consequently when such a sample is switched, the hysteresis loop constricts with a lowering of the coercive field, and depending on the extent of the depletion layer, the remnant polarisation as well \cite{Zubko2006, Wang2012, Baudry2005,Warren1994}. If the distribution of charges in the sample is asymmetric, the resulting internal electric field leads to a preferred polarisation direction, thereby shifting the hysteresis loop along the electric field axis; this phenomena is called imprint \cite{Wang2012,Misirlioglu2011effect}. 

Using a scanning probe microscope (SPM) tip to apply an electric field over a small region in a (100) \bfo\ film, Vasudevan \textit{et al.} observed the formation of domains with polarisations that form a closed loop (i.e. head-to-tail configuration of domains) at low applied fields, and domains with polarisations directed towards a central point (head-to-head configuration) at very high fields. \cite{Vasudevan2011}. Similar patterns may also be formed within space charge regions in ferroelectric films due to the internal electric field induced by the charges. Head-to-tail and 180\textdegree\ domains, whose walls are charge-neutral, are the more stable configurations in a charge-free material, whereas charged walls between head-to-head or tail-to-tail domains may stabilise in the presence of space charge \cite{Park1998,Gureev2011,Wu2006,Zuo2014}. Usually in theoretical investigations concerning the latter type of domains, the models have flat interfaces between the domains \cite{Wu2006, Zuo2014, Sluka2012, Eliseev2011}. Other studies, both experimental \cite{Randall1987,Han2014,Abplanalp1998} and theoretical \cite{Misirlioglu2012}, report the formation of zigzagged domain walls. It has been suggested that the formation of such interfaces allows the electrostatic energy to be reduced, as locally at the domain walls the polarisations are rotated into head-to-tail configurations \cite{Randall1987}. Misirlioglu \textit{et al.} find a critical thickness of the space charge layer above which this type of domain wall develops in their simulated BaTiO\sub{3} films, and below which the film remains as a single domain \cite{Misirlioglu2012}.

It can therefore be deduced that an approach to engineer complex domain structures is to manipulate the space charge regions in ferroelectric materials.

The thickness of space charge layers in thin films is determined by the built-in voltage across the layer. For a given dopant concentration, the built-in voltage is a property that depends on the material of the electrode. The layer thickness, which determines whether the film is fully or partially depleted, is thus dependent on the selection of dopant concentration and the electrode in experimental set-ups. When simulating a charged layer within a film, the layer thickness and dopant concentration may be chosen as variables instead; this would be equivalent to specifying the built-in voltage across the layer.

In investigating the properties of ferroelectric materials, the phase-field method has been shown to be a powerful technique and has been widely used \cite{Chen2008}. A real-space approach with this technique takes into account long-range elastic and electrostatic interactions in the material of study, and is convenient for nanoscale systems in which free surfaces are important since it is straightforward to apply the boundary conditions \cite{Yang2012}. This method has been applied in simulating the domain patterns in free-standing nanostructures of tetragonal lead zirconate-titanate \cite{Ng2012domain}. 

Most of the studies of space charge effects in ferroelectric materials have been devoted to materials such as  BaTiO\sub{3} \cite{Ng2012, Zuo2014, Sluka2012, Abplanalp1998, Misirlioglu2012}, PbTiO\sub{3} \cite{Wang2012, Park1998, Wu2006} and PbZr\sub{1-x}Ti\sub{x}O\sub{3} (PZT) for $x> 0.47$ \cite{Warren1994, Misirlioglu2011effect}, all of which adopt the tetragonal ferroelectric phase. These systems have 90\textdegree\ and 180\textdegree\ domain walls, with only the former producing a ferroelastic configuration. In these works, the external electric field is usually applied parallel to the polar direction. 

Few have investigated the effect of space charge on rhombohedral ferroelectric materials \cite{Vasudevan2011, Randall1987}. Unlike tetragonal ferroelectric materials, the rhombohedral phase has a larger number of ferroelectric variants with two possible ferroelastic domain patterns of 71\textdegree\ and 109\textdegree\ domain walls. Thus it can have more complex domain patterns \cite{Yin2000}, whose piezoelectric response may be enhanced especially when applying an electric field that is not along one of the polar directions \cite{Sluka2012, Fu2000, Wada1999, Bellaiche2001, Ahluwalia2005, Wang2007, Liang2010}. 

A rhombohedral ferroelectric material, \bfo, has been attracting much interest recently because of its multiferroic \cite{Ramesh2007, Catalan2009} and photoelectric \cite{Choi2009, Yang2009} properties. It has potential applications in memory storage, solar energy storage, sensors and telecommunications devices \cite{Catalan2009, Guo2013, Gao2007, Nuraje2013}.

Therefore in this work, we select \bfo\ with polarisations along the $\langle 111\rangle $ directions (with reference to the pseudocubic lattice) as our material of study and examine the domain structures that form in [001] thin films under the influence of space charge. We also investigate the effective polarisation-electric field behaviour when applying an electric field in the [001] direction. 

Assuming static charges on the timescale of polarisation switching  \cite{Ng2012} for the case where the space charge region is generated due to the presence of an electrode/film interface, what we find is a wealth of domain patterns that varies with the strength of the electric field induced by the charges.

Calculating the response in the direction parallel to the applied field, we note an optimum charge density with minimum charged layer thickness that produces a domain pattern with a wavy interface whose mobility is higher than the others. Our findings would be useful for applications that require materials with high piezoelectric response and small hystereses.

\section{Computational details}

The dynamics of the polarisation order parameter components $P_x$, $P_y$ and $P_z$ evolving with time $t$ is given by the time-dependent Ginzburg-Landau equation as
\begin{equation}
\frac{\partial P_i}{\partial t}=-\Gamma \left[\frac{\delta F}{\delta P_i }-E_i\right]
\label{ginzburglandau}
\end{equation}
where $\Gamma$ is a kinetic coefficient related to the domain wall mobility, $F$ is the free energy of the ferroelectric material and $E_i$ is the component $i$ of the electric field $\mathbf{E}$ in the sample. $F$ consists of the Landau free energy $F_l$, the gradient free energy associated with the domain walls $F_g$, the elastic free energy $F_e$ and the electrostatic free energy $F_{el}$:
\begin{equation}
F= F_l+ F_g+ F_e + F_{el}
\end{equation}
We express $F_l$ in terms of $P_x$, $P_y$, and $P_z$ up to the fourth order:
\begin{align}
F_l=\int d\mathbf{r} & \Big[ \alpha_1 \left(  P_x^2 + P_y^2 + P_z^2 \right) 
+\alpha_{11} \left( P_x^4 + P_y^4 + P_z^4 \right) \nonumber\\ 
& + \alpha_{12} \left( P_x^2 P_y^2 + P_x^2 P_z^2 + P_y^2 P_z^2 \right) \Big]
\end{align}
where $\alpha_1$, $\alpha_{11}$ and $\alpha_{12}$ are material parameters. $F_g$ can be expressed as:
\begin{equation}
F_g = \int d\mathbf{r} \left( \frac{K}{2} \left[  \left( \nabla P_x \right) ^2 + \left( \nabla P_y \right) ^2 + \left( \nabla P_z \right) ^2 \right] \right)
\end{equation}
where $K$ is a gradient coefficient, while the expression for $F_e$ is as below:
\begin{equation}
\begin{split}
F_e = & \int d\mathbf{r} \bigg\{
\frac{C_{11}}{2} \left[ \left( \varepsilon_{xx} - \varepsilon^0_{xx} \right) ^2 + \left( \varepsilon_{yy} - \varepsilon^0_{yy} \right) ^2 \right. \\ 
&\left. + \left( \varepsilon_{zz} - \varepsilon^0_{zz} \right) ^2 \right] \\ 
&+ C_{12} \left[ \left( \varepsilon_{xx} - \varepsilon^0_{xx} \right) \left( \varepsilon_{yy} - \varepsilon^0_{yy} \right) \right. \\ 
&\left. + \left( \varepsilon_{xx} - \varepsilon^0_{xx} \right) \left( \varepsilon_{zz} - \varepsilon^0_{zz} \right) \right. \\ 
&\left. + \left( \varepsilon_{yy} - \varepsilon^0_{yy} \right) \left( \varepsilon_{zz} - \varepsilon^0_{zz} \right) \right] \\ 
&+ 2C_{44} \left[ \left( \varepsilon_{xy} - \varepsilon^0_{xy} \right) ^2 + \left( \varepsilon_{xz} - \varepsilon^0_{xz} \right) ^2 \right. \\ 
&\left. + \left( \varepsilon_{yz} - \varepsilon^0_{yz} \right) ^2 \right] \bigg\}
\end{split}
\end{equation}
$C_{11}$, $C_{12}$ and $C_{44}$ are material constants. $\varepsilon_{ij}$ are the strain tensors having the usual definition in terms of displacements $u_i$
\begin{equation}
\varepsilon_{ij} = \frac{1}{2} \left( \frac{\partial u_i}{\partial x_j} + \frac{\partial u_j}{\partial x_i} \right)
\end{equation}
and the electrostrictive strain tensors $\varepsilon^0_{ij}$ have the following expressions:
\begin{align}
\varepsilon^0_{ii} = & Q_{11} P^2_i + Q_{12} \left( P^2_j + P^2_k \right) \\
\text{and } \varepsilon^0_{ij} = & Q_{44} P_i P_j \text{ for } i\neq j
\end{align}
with $Q_{11}$, $Q_{12}$ and $Q_{44}$ being material-specific electrostrictive constants.

We determine the electric field $\mathbf{E}$ in the sample from the electrostatic potential $\phi$ using the relationship
\begin{equation}
\mathbf{E} = - \mathbf{\nabla} \phi
\end{equation}
as well as Gauss's law relating the electric displacement \textbf{D}, \textbf{E} and polarisation \textbf{P}:
\begin{equation}
\mathbf{\nabla} \cdot \mathbf{D} = \mathbf{\nabla} \cdot \left( \epsilon_0 \epsilon_{br} \mathbf{E} + \mathbf{P} \right)
= - \epsilon_0 \epsilon_{br} \nabla ^2 \phi + \mathbf{\nabla} \cdot \mathbf{P} = \rho \left( \mathbf{r} \right)
\end{equation}
$\epsilon_0$ is the permittivity of free space, $\epsilon_{br}=10$ is a background permittivity \cite{Hlinka2006} and $\rho \left( \mathbf{r} \right)$ we have set as 
\begin{equation}
\rho \left( z \right) = 
\begin{cases}
    \rho _0 ,& L - z < w \\
    0, & \text{otherwise}
\end{cases}
\end{equation}
where $L$ is the thickness of the film and $w$ is the thickness of the space charge layer. We assume $\rho_0 = q N_D$ where $q$ is the charge of a carrier and $N_D$ is the defect (or carrier) density. In our simulations, we arbitrarily select $q=-e$ where $e$ is the elementary charge. We now have $F_{el}$, defined as 
\begin{equation}
F_{el} = \int d\mathbf{r} \left( \frac{\varepsilon_0\varepsilon_{br}}{2} E_i \cdot E_i \right)
\end{equation}

The dissipative force balance equations give the displacement field dynamics:
\begin{equation}
\bar{ \rho} \frac{\partial ^2 u_i}{\partial t^2} = \frac{\partial \sigma_{ij} }{\partial x_j} + \eta \nabla ^2 \frac{\partial u_i}{\partial t}
\label{mechanical}
\end{equation}
where $\bar{\rho}$ is the density and $\eta$ is a viscous damping term. The latter functions to drive the system to mechanical equilibrium such that $\frac{\partial \sigma_{ij} }{\partial x_j} = 0$. The stress tensors $\sigma_{ij}$ are obtained as
\begin{align}
\sigma_{ij} = \frac{\delta F}{\delta \varepsilon_{ij} }
\label{final}
\end{align}

We consider a (001) epitaxial \bfo\ film on top of a substrate, with an electrode deposited at the top surface of the film, and simulate the dynamics of the polarisations by numerically solving the model described in equations \ref{ginzburglandau} to \ref{final} using finite differences. Initialising $P_x$, $P_y$ and $P_z$ with small, random fluctuations about zero to simulate the paraelectric initial conditions, we examine the ferroelectric domain patterns that form in fully and partially depleted films. We employ a $128 \text{nm} \times 128\text{nm} \times 32 \text{nm}$ simulation cell with periodic boundary conditions for \textbf{P}, \textbf{u} and $\phi$ in the lateral directions. The top surface is stress-free with $\sigma_{ij} \cdot n = 0$, where $n$ is the surface normal, and the bottom surface is clamped ($u_x = u_y = u_z = 0$). The film is also laterally clamped as we have periodic boundary conditions on $u_x$ and $u_y$ laterally. Assuming that the top and bottom interface charges compensate exactly the respective bound interface charges, we impose $\frac{\partial P_z}{\partial z} = 0$ at $z = 0$ and $z = L = 32$ nm. Rescaling all lengths as $\mathbf{r} = \mathbf{r}^* \delta $, we choose $\delta=1$ nm as our grid spacing. We rescale the polarisation as $\mathbf{P} = P_0 \mathbf{P}^*$ where $P_0=0.9$ Cm$^{-2}$ is the spontaneous polarisation of \bfo, the strains as $\varepsilon_{ij} = \varepsilon_0 \varepsilon_{ij}^*$ where $\varepsilon_0 = Q_{11} P_0^2$, and the time as $t=t_0t^*$ where $t_0 = (\Gamma \lvert \alpha_1 \rvert ) ^{-1}$. All other coefficients are consequently rescaled in the same way as reported in reference \citenum{Ng2012domain}, and we use their values for rescaled coefficients of $\Gamma$, $K$, $\eta$ and $\bar{\rho}$. The material specific parameters are taken from reference \citenum{Winchester2011} and listed in Table \ref{param}. We model films with values of $N_D$ between $10^{26}$ $\text{m}^{-3}$ to $10^{27}$ $ \text{m}^{-3}$, which are within the range of those for heavily doped \bfo\ films \cite{Yang2009electric, Hung2014}. The charges are uniformly distributed within space charge layers parallel and adjacent to the film/electrode interface with thicknesses $w$ of 0, 5, 10, 15 and 32 nm, with the latter thickness corresponding to the fully depleted case.

\begin{table}
\caption{The material constants for rhombohedral \bfo\ taken from reference \citenum{Winchester2011}. $T$ is the temperature in Kelvin. \label{param}}
\begin{tabular}{c c}
\hline
$\alpha_{1}$ & $4.9(T-1103)\times 10^5$ V m/C \\ 
$\alpha_{11}$ & $6.5 \times 10^8$ V m$^5$/C$^3$ \\ 
$\alpha_{12}$ & $1.0 \times 10^8$ V m$^5$/C$^3$ \\ 
$C_{11}$ & $3.00 \times 10^{11}$ N/m$^2$ \\ 
$C_{12}$ & $1.62 \times 10^{11}$ N$^9$/m$^2$ \\ 
$C_{44}$ & $0.691 \times 10^{11}$ N$^9$/m$^2$ \\ 
$Q_{11}$ & 0.032 m$^4$/C$^2$ \\ 
$Q_{12}$ & -0.016 m$^4$/C$^2$ \\ 
$Q_{44}$ & 0.02 m$^4$/C$^2$ \\ 
\hline
\end{tabular}
\end{table}

\section{Results and discussions}

\subsection{The charge-free \bfo\ thin film}

\begin{figure*}
\begin{minipage}[b]{0.33\linewidth}
\subfigure[]{\includegraphics[width=\textwidth]{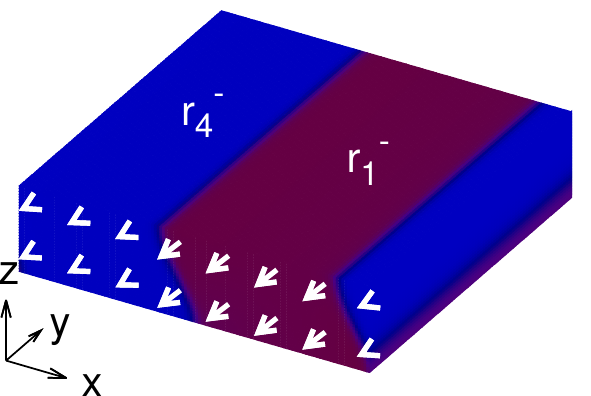}\label{uncharged_domain}}\\
\subfigure[]{\includegraphics[width=\textwidth]{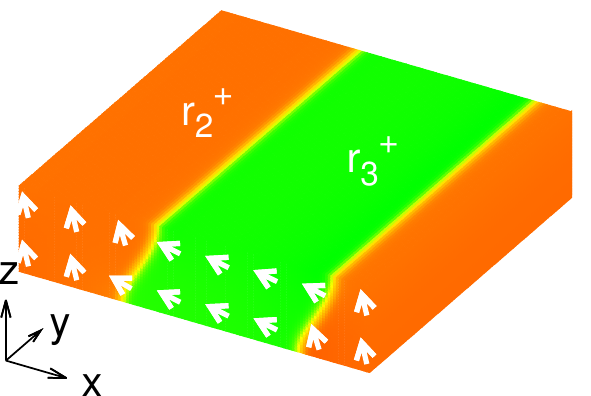}\label{positive_remnant_domain}}
\end{minipage}
\begin{minipage}[b]{0.66\linewidth}
\subfigure[]{\includegraphics[width=\textwidth]{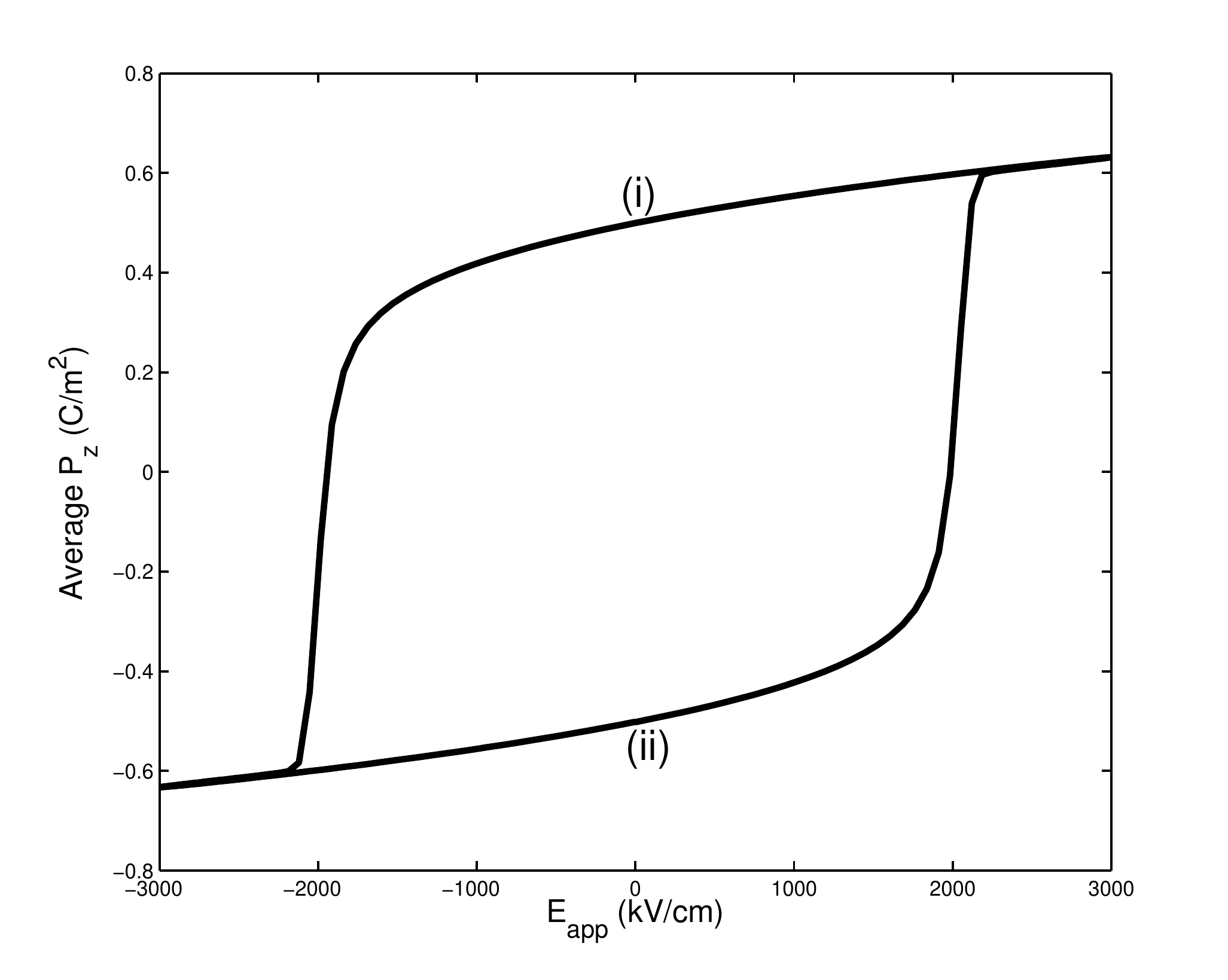}\label{uncharged_loop}}
\end{minipage}
\caption{(a) The domain pattern of \bfo\ thin film without presence of charge, with the direction of polarisation indicated for $y=0$ plane. The colour within the film distinguishes between the domains. This domain pattern also corresponds to the point labelled (ii) in (c). (b) The domain pattern of \bfo\ thin film without presence of charge corresponding to the remnant state labelled (i) in (c). (c) The simulated $P_z-E_{app}$ hysteresis loop for the film. \label{uncharged}}
\end{figure*}

Bulk \bfo\ in the rhombohedral phase has eight possible polarisation variants along the $\langle 111\rangle $ directions, which are commonly denoted as $r_1^+ = \left[ 1 1 1 \right]$, $r_2^+ = \left[ \bar{1}  1 1 \right]$, $r_3^+ = \left[ \bar{1} \bar{1}  1 \right]$, $r_4^+ = \left[ 1 \bar{1}  1 \right]$, $r_1^- = \left[ \bar{1} \bar{1} \bar{1} \right]$, $r_2^- = \left[ 1 \bar{1} \bar{1} \right]$, $r_3^- = \left[ 1 1 \bar{1} \right]$, $r_4^- = \left[ \bar{1} 1 \bar{1} \right]$ (with reference to the pseudocubic lattice). Without any presence of charges in the film, our \bfo\ thin film evolves from the paraelectric state to form 71\textdegree\ twinned domains with domain walls at an angle greater than 45\textdegree\ to the film/substrate interface, consistent with results reported by Zhang \textit{et al.} using nearly similar material constants in their phase-field model \cite{Zhang2008}. Figure \ref{uncharged_domain} shows $r_1^-/r_4^-$ domains that form in some of our simulations when steady state is reached. With different random seeds to initialise the polarisations in the paraelectric state, we have observed other twinned domain formations with \{101\}-type walls as suggested by Streiffer \textit{et al.} for films with rhombohedral lattices \cite{Streiffer1998} such as $r_3^+/r_4^+$, $r_2^+/r_3^+$ and $r_1^-/r_2^-$, to name a few. They also noted that all domains in a film must have the same $P_z$ direction to maintain a charge-neutral film. In all of our simulated films, the domain walls deviate slightly from the ideal \{101\} orientation; Zhang \textit{et al.} have shown that this deviation is due to the competition between the domain wall energy that minimises when the domain wall is perpendicular to the film/substrate interface, and the elastic energy that favours the ideal orientation at 45\textdegree\ to the interface \cite{Zhang2008}.

We then apply a varying potential difference across the thickness of the film with $r_1^-/r_4^-$ domains to simulate the ferroelectric switching of the charge-free film, by varying the external electric field from 0 kV/cm to 3000 kV/cm, $-3000$ kV/cm and back to 0 kV/cm in $t^*=10^5$ timesteps. Computing the average $P_z$ as a function of the applied field $E_{app}$ we show the resulting curve in Figure \ref{uncharged_loop}. The $r_1^-/r_4^-$ domains are observed for the remnant state with $P_z < 0$, corresponding to the point labelled as (ii) in Figure \ref{uncharged_loop}. These domains switch fully to $r_3^+/r_2^+$ domains upon forward bias, and back to $r_1^-/r_4^-$ domains with reverse bias. The $r_3^+/r_2^+$ domain pattern, illustrated in Figure \ref{positive_remnant_domain}, is stable at the remnant state with $P_z > 0$ which is labelled as (i) in Figure \ref{uncharged_loop}. The $P_x$ and $P_y$ components of the polarisation in the film remain unchanged during the switching as we expect, with only the $P_z$ component switching, since we are applying the electric field in the [001] direction. As we switch between the two remnant states, there is no long-range migration of the domain wall; only the inclination of the wall is altered.

\subsection{Domain formation in fully and partially depleted films}

\begin{figure*}
\subfigure[$N_D=3 \times 10^{26}$ m$^{-3}$]{\includegraphics[width=0.3\textwidth]{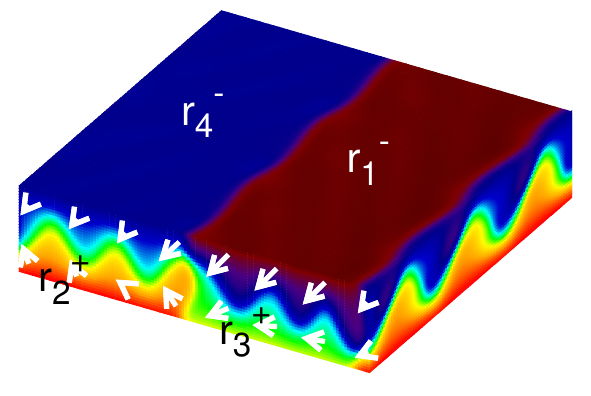}\label{domain_rho0_05}} \quad
\subfigure[$N_D=6 \times 10^{26}$ m$^{-3}$]{\includegraphics[width=0.3\textwidth]{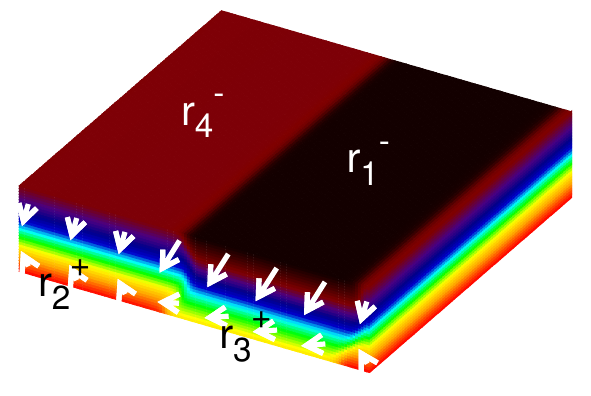}\label{domain_rho0_1}} \quad
\subfigure[$N_D=1.1 \times 10^{27}$ m$^{-3}$]{\includegraphics[width=0.3\textwidth]{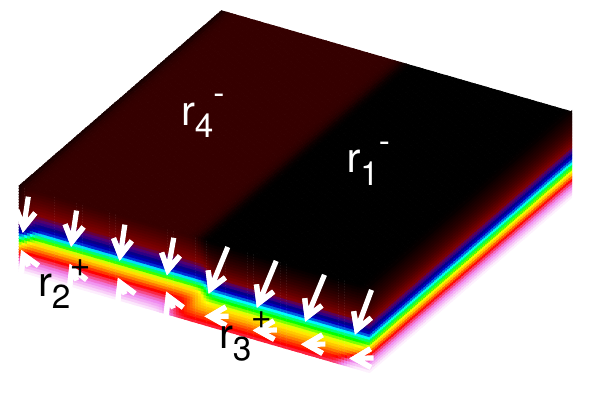}\label{domain_rho0_2}} 
\caption{Domain patterns in fully depleted films with $N_D$ values from $3 \times 10^{26}$ m$^{-3}$ to $1.1 \times 10^{27}$ m$^{-3}$. Lighter, brighter shades represent domains with more positive $P_z$ (i.e. $r_n^+$) and darker shades, those with more negative $P_z$ (i.e. $r_n^-$). \label{fullydepleteddomains}}
\end{figure*}

When we include uniformly distributed static charges throughout the film to model a fully depleted film, we find that 
instead of having only twinned $r_1^-/r_4^-$ domains in the film, the top of the film consists of a set of $[P_x,P_y,-P_z]$ and $[P_x,-P_y,-P_z]$ twinned domains and the bottom half, $[P_x,P_y,P_z]$ and $[P_x,-P_y,P_z]$ domains. Figure \ref{fullydepleteddomains} illustrates these domain configurations forming in fully depleted films for different values of $N_D$, clearly demonstrating that the geometry of the domain wall between these sets of twins depends on the magnitude of the charge density. The film with a lower value of $N_D=3 \times 10^{26}$ m$^{-3}$ has a wavy domain wall in the lateral directions of the film separating the sets of twinned domains  that meet in a head-to-head configuration (Figure \ref{domain_rho0_05}). For high values of $N_D$, this wall has a planar geometry (Figures \ref{domain_rho0_1} and \ref{domain_rho0_2}). 

\begin{figure}
\includegraphics[width=\columnwidth]{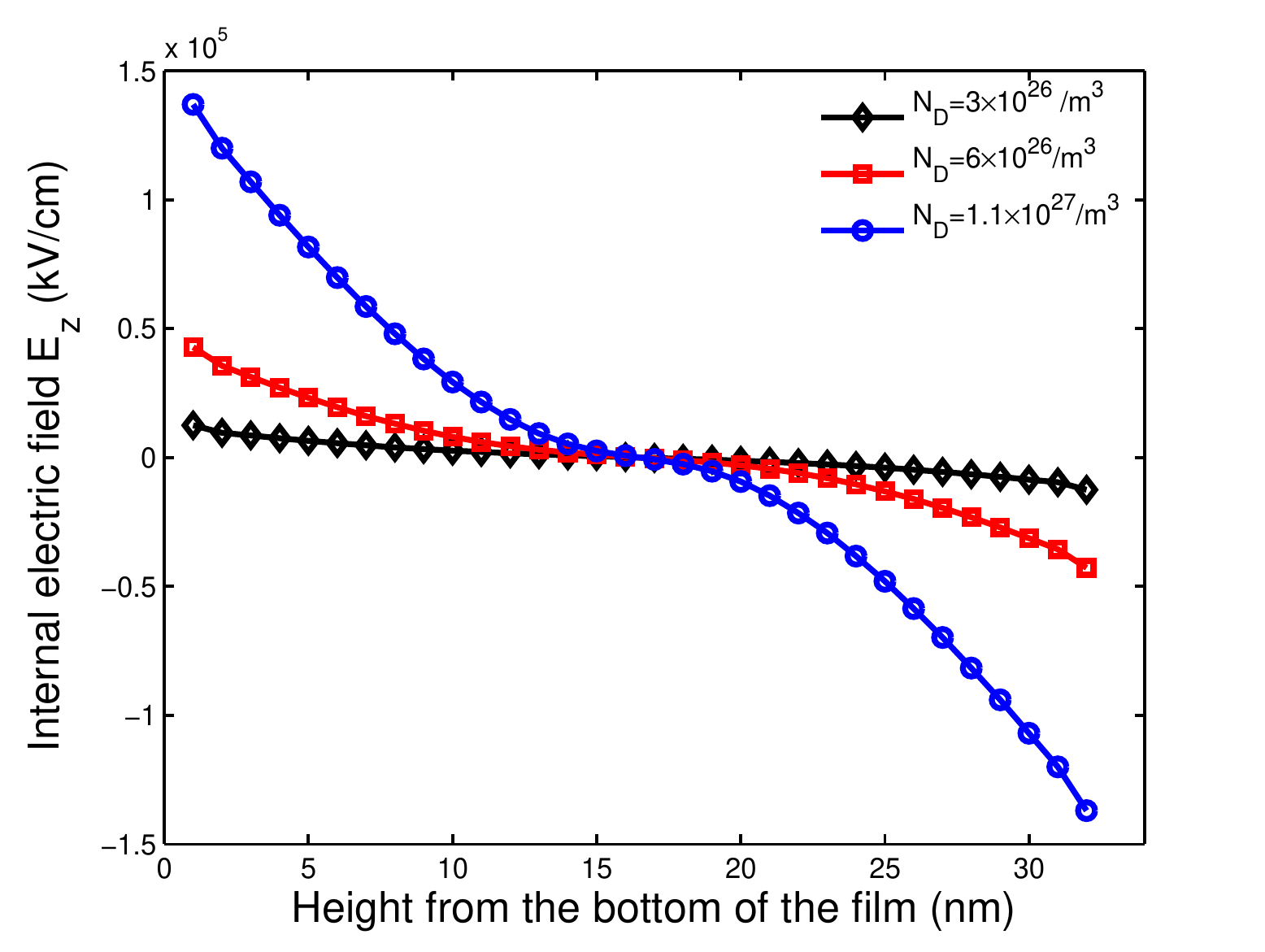}
\caption{The profile of the internal field $E_z$ (in the paraelectric phase) induced by uniformly distributed charges in fully depleted films with $N_D$ values ranging from $3 \times 10^{26}$ m$^{-3}$ to $1.1 \times 10^{27}$ m$^{-3}$ across the thickness of the films. \label{electricfield}}
\end{figure}

To understand the formation of the domain configurations described above, we examine the internal electric field induced by the charges. A uniform charge density produces a radial electric field inside a charged body. For the thin film geometry considered in our work, the field varies only in the $z$ direction. This internal field may influence the domain pattern that forms when the sample is cooled from the paraelectric phase to below the ferroelectric transition temperature. Since the field is already present in the paraelectric phase, we perform additional simulations in the paraelectric phase for the same values of $N_D$ as before to determine the variation of the internal electric field within the films. As shown in Figure \ref{electricfield}, the internal electric field generated due to the uniform distribution of charges in a film is equal in magnitude but opposite in direction in the top and bottom halves of the film. Since we have introduced negative charges in the film, the electric field in the bottom half is positive while that in the top half is negative. The magnitude of the field is maximum at the top and bottom surfaces and decreases to zero towards the middle of the film. 

The electric field will induce a rotation of the polarisations to align with the field locally if the magnitude of the field is large enough. For films with sufficiently high values of $N_D$ to generate large enough internal fields, the positive field in the bottom half of the charged region stabilises domains with $P_z > 0$ and the negative field in the top half, domains with $P_z <0$. In rhombohedral ferroelectrics, this means that only $r_1^+$, $r_2^+$, $r_3^+$ and $r_4^+$ may form in the bottom half and $r_1^-$, $r_2^-$, $r_3^-$ and $r_4^-$ in the top half. In our simulations, two of the four possible domains in each half are formed, where the domains in one half must each be twins of those in the other half through the (001) twin plane, giving rise to the four-domain pattern observed in Figure \ref{fullydepleteddomains}. Thus it is the inhomogeneity in the internal electric field that is responsible for the formation of multiple domains within the charged region.

\begin{figure}
\subfigure[]{\includegraphics[scale=0.32,angle=270,bb=30 320 200 380]{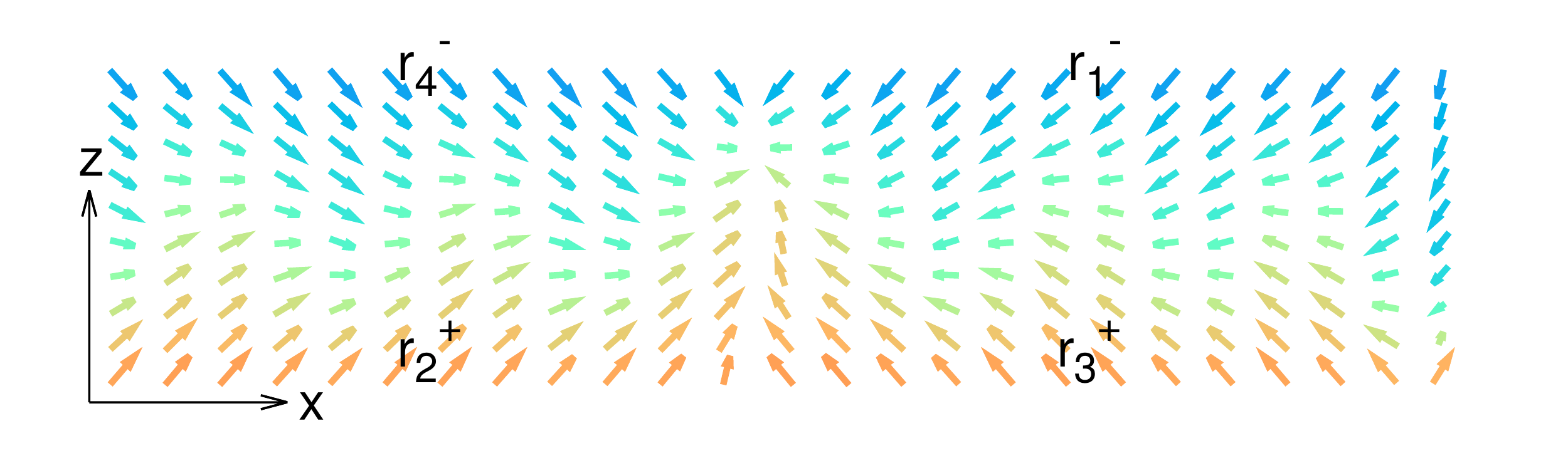}\label{vector_32rho0_05}} \\
\subfigure[]{\includegraphics[scale=0.32,angle=270,bb=30 320 200 380]{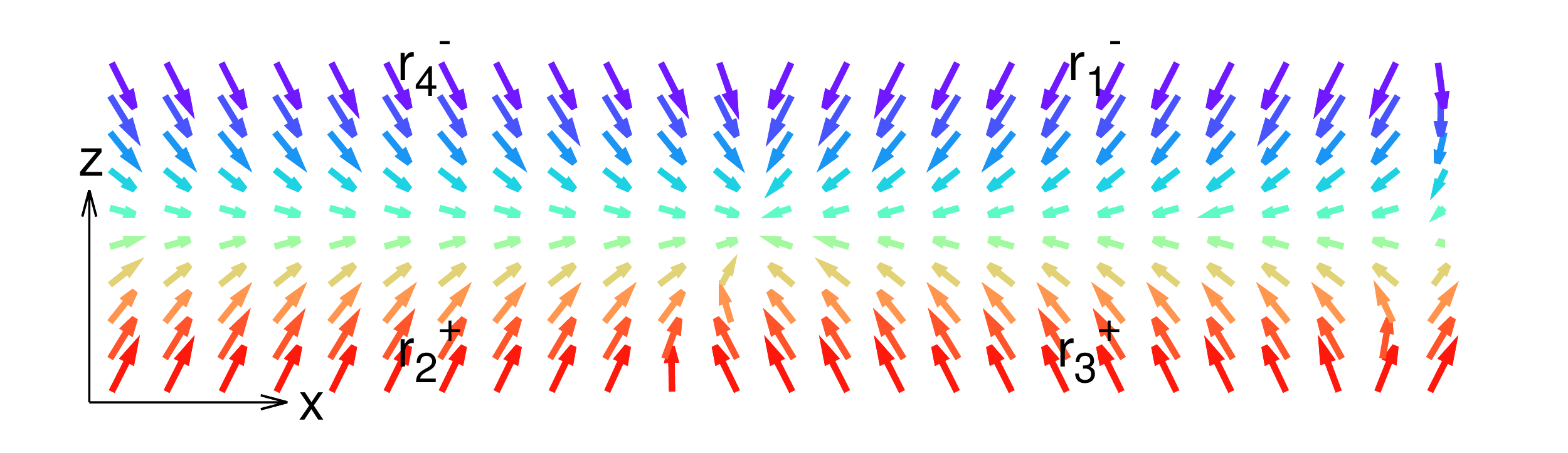}\label{vector_32rho0_1}}  \\
\subfigure[]{\includegraphics[scale=0.32,angle=270,bb=30 320 200 380]{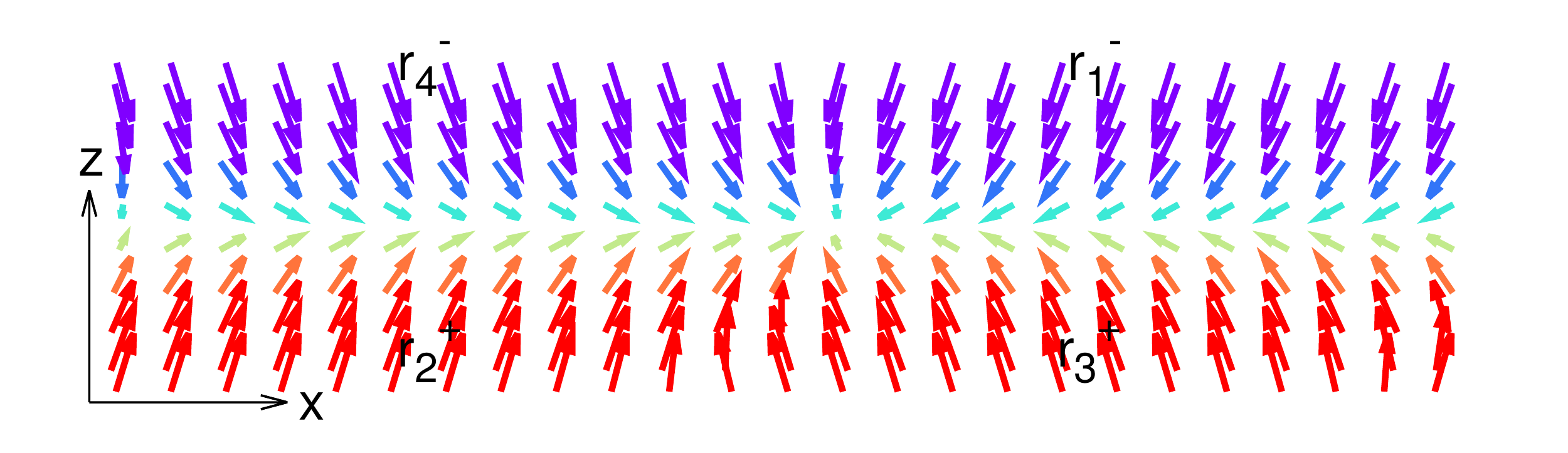}\label{vector_32rho0_2}}  
\caption{Polarisation vector maps of $P_z-P_y$ in an $x-z$ plane of fully depleted films with $N_D$ values of (a) $3 \times 10^{26}$ m$^{-3}$, (b) $6 \times 10^{26}$ m$^{-3}$ and (c) $1.1 \times 10^{27}$ m$^{-3}$. Domains $r_4^-$ (top left of film), $r_2^+$ (bottom left of film), $r_1^-$ (top right of film) and $r_3^+$ (bottom right of film) are indicated. These maps correspond to the domains illustrated in Figure \ref{fullydepleteddomains}. \label{Pvector} }
\end{figure}

Comparing the variation of the internal field for films with different values of $N_D$ (Figure \ref{electricfield}), we observe that the magnitude of the internal field within the charged region increases as $N_D$ increases. For low values of $N_D$, the small magnitude of the internal field is insufficient to produce a pattern of domains different from that in the charge-free film. For depleted films with intermediate values of $N_D$, the magnitude of the internal field which is larger in the near-surface regions, is sufficient to induce the formation of domains with polarisations that align with the field in these regions. Towards the middle of the film, however, the magnitude of the field is too small to force the polarisations to rotate with the field. In this case, the polarisations rotate into a head-to-tail configuration locally near the domain wall, accommodating the constraint imposed by the induced domains in the near-surface regions and at the same time minimising the electrostatic energy. Consequently the interface takes on a wavy geometry, as illustrated in Figure \ref{vector_32rho0_05} showing the $P_z-P_y$ vectors in an $x-z$ plane for the fully depleted film with $N_D=3 \times 10^{26}$ m$^{-3}$.

When the value of $N_D$ is high, the magnitude of the electric field increases more rapidly from the middle of the film. The field is therefore large enough in the middle of the film to induce a head-to-head domain configuration with a planar interface (Figures \ref{Pvector}(b) and (c)). 
It also results in a rotation of the dipoles from pure rhombohedral orientations towards the [001] and $\left[ 0 0 \bar 1 \right]$ directions, such that for the fully depleted film with $N_D=1.1 \times 10^{27}$ m$^{-3}$, the domains approach the tetragonal phase. The increasing $P_z$ component in the near-surface regions of films with increasing $N_D$ can be observed from the $P_z-P_y$ vector maps in Figure \ref{Pvector}, and we also represent the degree of the polarisation rotation with the changing shades within the domains in Figure \ref{fullydepleteddomains}.

\begin{figure*}
\subfigure[$w=5$ nm, $N_D=3 \times 10^{26}$ m$^{-3}$] {\includegraphics[width=0.3\textwidth]{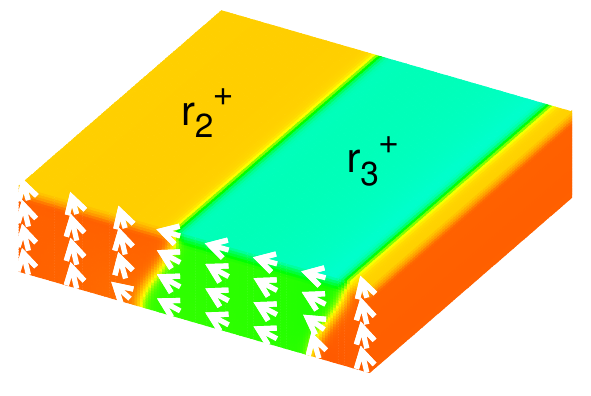}} \quad
\subfigure[$w=5$ nm, $N_D=6 \times 10^{26}$ m$^{-3}$]{\includegraphics[width=0.3\textwidth]{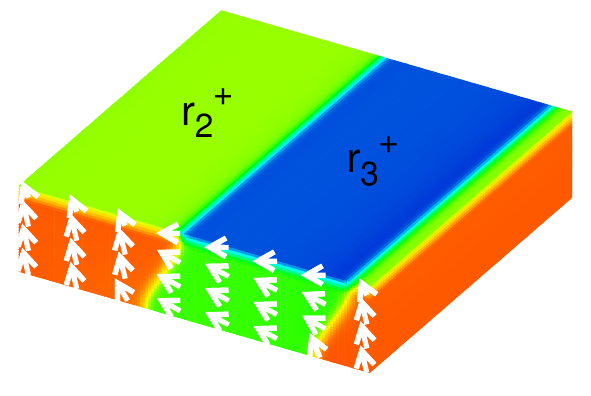}} \quad
\subfigure[$w=5$ nm, $N_D=1.1 \times 10^{27}$ m$^{-3}$]{\includegraphics[width=0.3\textwidth]{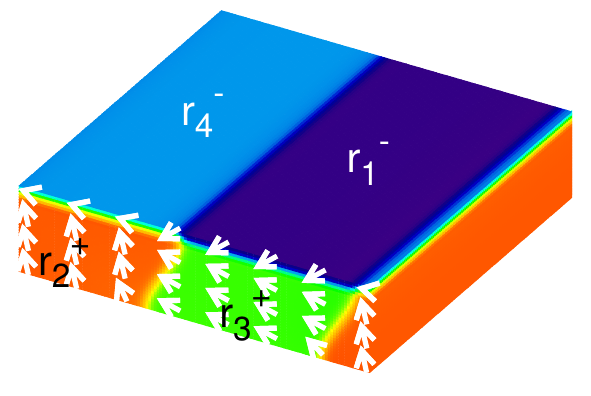}} \\
\subfigure[$w=10$ nm, $N_D=3 \times 10^{26}$ m$^{-3}$]{\includegraphics[width=0.3\textwidth]{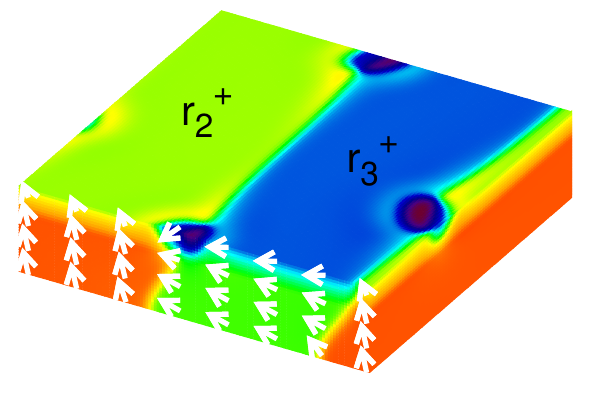}} \quad
\subfigure[$w=10$ nm, $N_D=6 \times 10^{26}$ m$^{-3}$]{\includegraphics[width=0.3\textwidth]{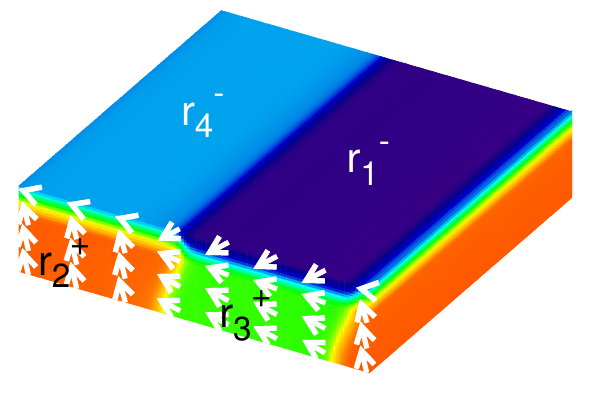}} \quad
\subfigure[$w=10$ nm, $N_D=1.1 \times 10^{27}$ m$^{-3}$]{\includegraphics[width=0.3\textwidth]{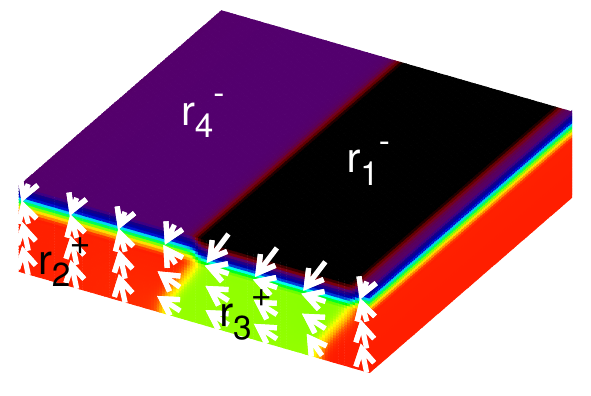}} \\
\subfigure[$w=15$ nm, $N_D=3 \times 10^{26}$ m$^{-3}$]{\includegraphics[width=0.3\textwidth]{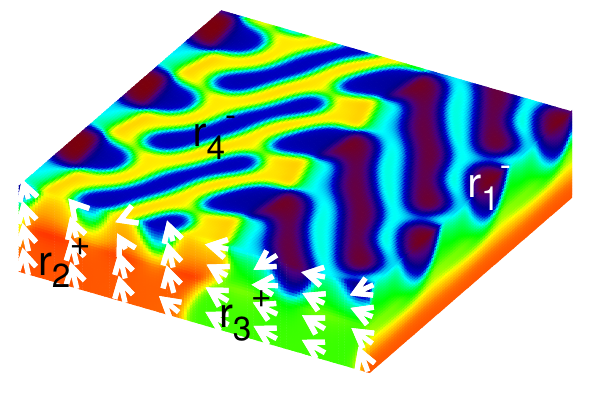}} \quad
\subfigure[$w=15$ nm, $N_D=6 \times 10^{26}$ m$^{-3}$]{\includegraphics[width=0.3\textwidth]{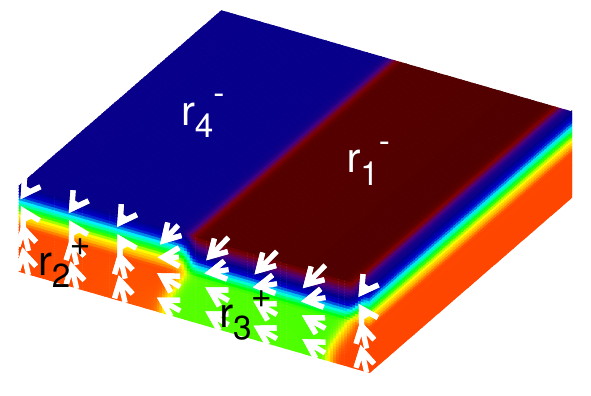}} \quad
\subfigure[$w=15$ nm, $N_D=1.1 \times 10^{27}$ m$^{-3}$]{\includegraphics[width=0.3\textwidth]{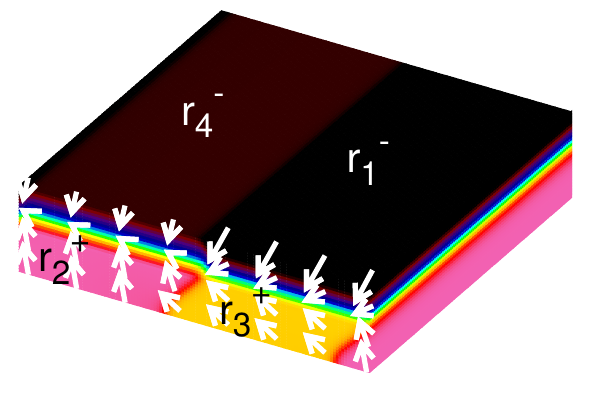}} \\
\caption{Domain patterns in partially depleted films with depletion layer thicknesses from 5 nm to 15 nm, and $N_D$ values from $3 \times 10^{26}$ m$^{-3}$ to $1.1 \times 10^{27}$ m$^{-3}$. \label{domains}}
\end{figure*}

In partially depleted films, the presence of a charged layer shows a similar effect. For films with uniformly distributed charges of $N_D=3\times 10^{26}$ $\text{m}^{-3}$ and $N_D=6\times 10^{26}$ $\text{m}^{-3}$ within charged layers of width $w=5$ nm at the top surface of the films, the domain patterns resemble that of a charge-free film with twin $r_3^+/r_2^+$ domains as shown in Figures \ref{domains}(a) and (b), except for a slight decrease in $P_z$ near the top of the layer (indicated by the different shades in this region). When we increase $N_D$ to $1.1\times 10^{27}$ $\text{m}^{-3}$ for the same value of $w$, we observe the formation of the $r_1^-/r_4^-$ twinned set of domains in the top half of the charged layer and the $r_3^+/r_2^+$ set in the bottom half of the charged layer, with the latter set extending into the charge-free region (Figure \ref{domains}(c)). 

We obtain a similar trend for films with $w=10$ nm, but with the appearance of the double set of twinned domains from lower values of $N_D$. Small regions of $r_1^-/r_4^-$ domains already form at the domain walls in the film with $N_D=3\times 10^{26}$ $\text{m}^{-3}$, represented as the dark-shaded regions in Figure \ref{domains}(d). Increasing $w$ to 15 nm with $N_D=3\times 10^{26}$ $\text{m}^{-3}$ leads to the appearance of such domains in a larger number (Figure \ref{domains}(g)), such that the domain pattern begins to resemble that of the fully depleted film having the same $N_D$ with the wavy domain interface (Figure \ref{domain_rho0_05}). The $r_3^+/r_2^+$ set of domains from the lower half of the film extends up to the top surface of the film in some regions, as $w=15$ nm is not sufficiently large to allow the formation of a continuous wavy interface. For films with this value of $w$, and $N_D$ values of $6 \times 10^{26}$ $\text{m} ^{-3}$ and $1.1 \times 10^{27}$ $\text{m} ^{-3}$, the two sets of twinned domains form with planar interfaces, with rotation of the polarisations approaching the tetragonal phase as before even within the charge-free regions. 

\begin{figure}
\includegraphics[width=0.45\textwidth]{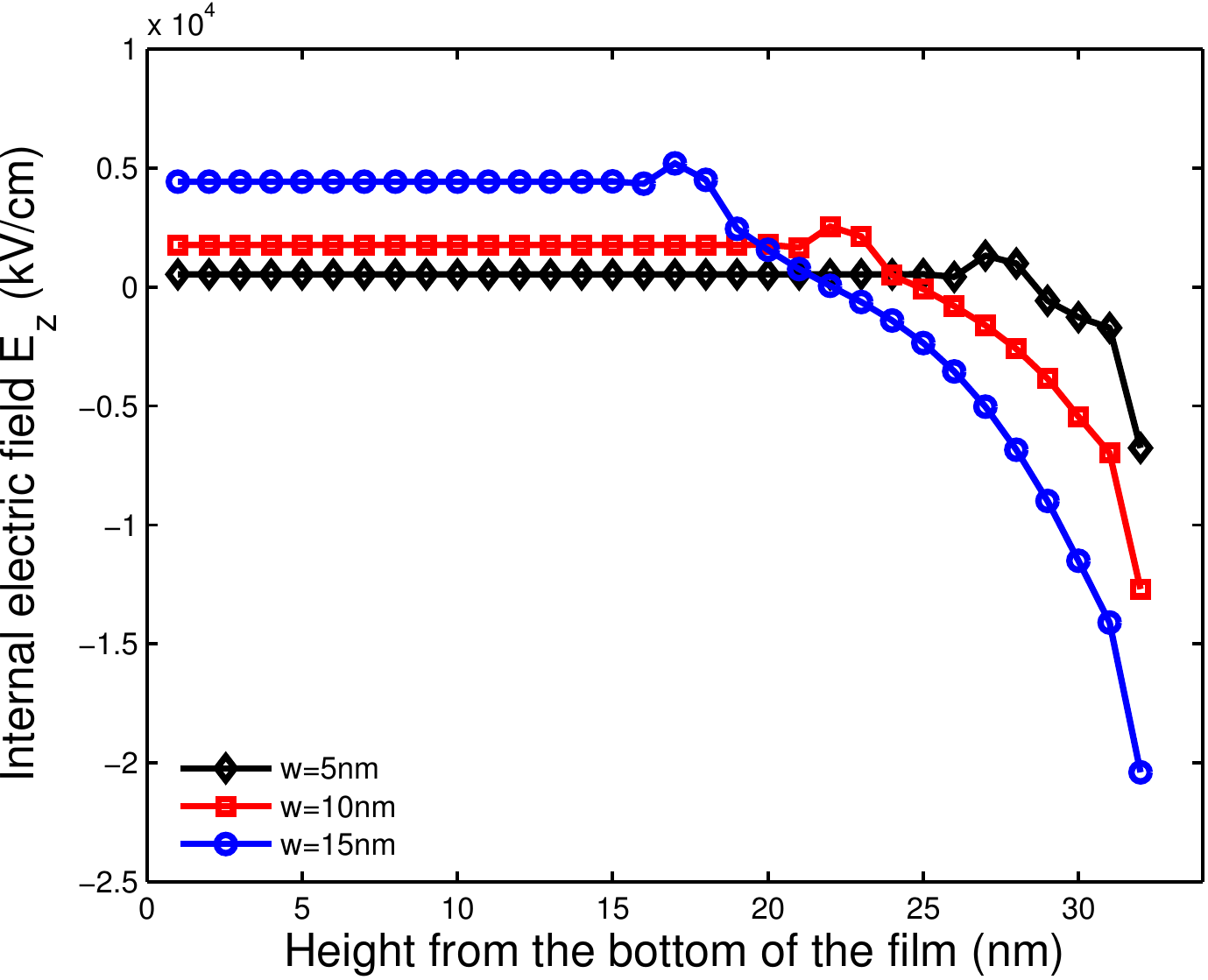}
\caption{The $E_z$ profile across the thickness of partially depleted films with $N_D=6 \times 10^{26}$ m$^{-3}$, for values of $w$ from 5 nm to 15 nm. \label{efield_rho0_1}}
\end{figure}

Whether the set of $r_1^-/r_4^-$ domains forms within the charged layer of a partially depleted film depends on the strength of the internal electric field, determined by the values of $w$ and $N_D$. The field within the charged region, whose profile is shown in Figure \ref{efield_rho0_1} for films with  $N_D=6 \times 10^{26}$ m$^{-3}$, does not induce the formation of these domains in the top half of the charged layer if the magnitude of the field is too small. As the magnitude increases with increasing $w$ and $N_D$, the $r_1^-/r_4^-$ domains are observed near the top of charged layers with higher values of these parameters. Within the much larger charge-free region, the internal field is non-zero and positive, increasing in magnitude with increasing $N_D$ and $w$. Although the magnitude in this region is much smaller than the maximum magnitude attained at the top of the charged layer, it is sufficient to bias the formation of the $r_3^+/r_2^+$ domains within the charge-free region, with the domains approaching tetragonality with increasing magnitude of the internal field.

The appearance of the domain pattern with the wavy domain interface for $w \geq 15$ nm is consistent with the finding by Misirlioglu \textit{et al.}. From their 2-dimensional Ginzburg-Landau-Devonshire model of depleted tetragonal BaTiO\sub{3} films, they predict a critical thickness above which head-to-head domains with a zigzagged interface develop \cite{Misirlioglu2012}. We also observe a critical value of $w$ associated with the strength of the induced electric field for the appearance of the domain pattern with planar horizontal interface that we find in films with high $N_D$. For films with $N_D= 6 \times 10^{26}$ $\text{m} ^{-3}$, this configuration appears when $w \geq 10$ nm (Figure \ref{domains}(e)), while films with $N_D= 1 \times 10^{27}$ $\text{m} ^{-3}$ need only $w \geq 5$ nm for the same configuration to form (Figure \ref{domains}(c)).

There is in addition a critical value of $N_D$ beyond which the wavy domain wall transitions to a planar interface. We have observed this progression for the fully depleted films, as well as for the partially depleted films with $w=10$ nm (not shown here) and 15 nm (Figures \ref{domains}(g-i)). For a film with a thinner depletion layer of $w=5$ nm, the charged layer is too thin to allow the formation of a wavy wall. Thus the twinned domain pattern resembling the charge-free film directly transitions to the four-domain pattern with planar interface when $N_D \geq 1.1 \times 10^{27} \text{m} ^{-3}$. 

Various experimental studies have reported observations of domains in a head-to-head or tail-to-tail configuration with a wavy interface, one of which have recently been observed in a tetragonal PZT film \cite{Han2014}. The authors in that work attributed the domain shapes to the anisotropy of domain growth, and suggest opposite internal fields near the top and bottom surfaces of the film due to high oxygen vacancy concentration and electronic band bending at the film/substrate interface respectively. These are thought to induce domains with polarisations pointing down towards the middle of the film ($P_z<0$) near the top of the film, and domains with upwards polarisations ($P_z>0$) at the bottom of the film. Our results show that domains in a head-to-head configuration may be generated by a single negatively charged layer if the magnitude of the induced internal field is sufficiently high, with a residual field even within the charge-free regions. Furthermore, the geometry of the interface is dictated by the strength of the field. Having two oppositely charged layer in a film may produce the same domain configuration provided the induced field within the other positively charged layer is small enough in magnitude. If the magnitude is sufficiently high, additional domains in tail-to-tail configuration would result in the positively charged layer.

We observe that in both partially and fully depleted films that we have considered in our work, the induced internal fields are very large in magnitude. In principle, such large fields in the ferroelectric state may lead to charge transport which can modify the field and thus the domain patterns, particularly in films with large $N_D$. However, we expect the charge migration to be a slow process. For example, in a study of the ageing phenomenon in ferroelectrics, it has been shown that charge migration induced by a depolarisation field leads to a build-up of space charge over time scales ranging from 10$^5$ seconds to 10$^7$ seconds \cite{Genenko2008}. This is much slower than the time scales associated with domain reorientation, which is on the order of 10$^{-9}$ seconds. Thus we expect findings based on a model with static space charge to be valid for short times and fast switching frequencies. The fact that the resulting domain patterns, such as those with wavy interfaces, have indeed been observed in experiments for various systems \cite{Randall1987,Han2014,Abplanalp1998} gives us confidence that our observations will still hold in the long time limit, as the internal field may not be completely eliminated by mobile charges.

\subsection{Polarisation switching and response of the films}

When an external electric field is applied to a film with depleted layers, the internal field produced by the presence of charges may oppose the effect of the externally applied field. Within the thin films that we have considered in this work, we have shown that the internal field along the $z$-direction is positive in the lower regions of the charged layers but negative in the upper regions in the absence of external bias. Therefore one of these regions will always oppose an applied field parallel to the internal field. Consequently, the domains will be pinned, requiring prohibitively large applied fields to completely switch to domains having the same polarity in $z$. 

\begin{figure*}
\begin{minipage}[b]{0.49\linewidth}
\includegraphics[width=0.95\textwidth]{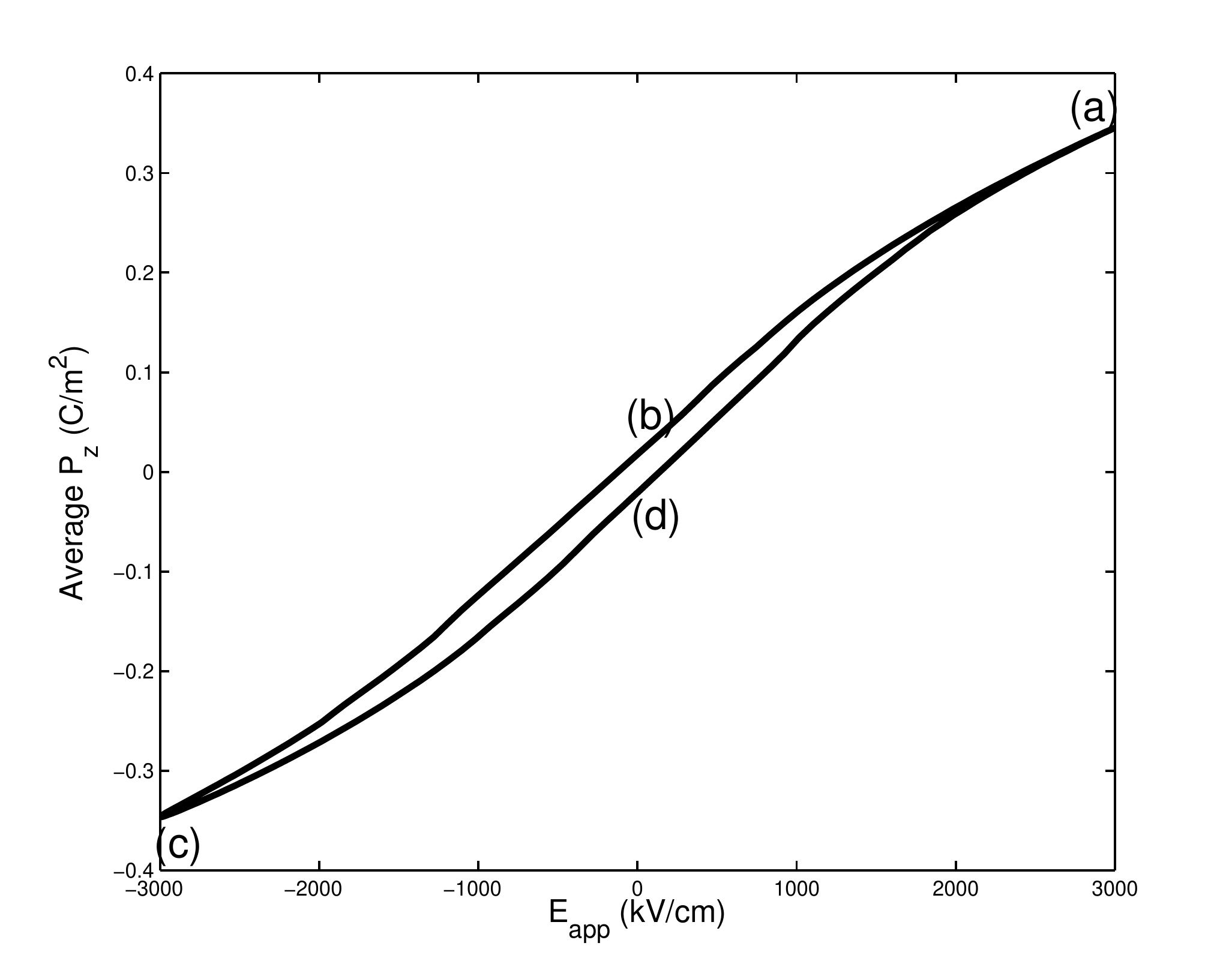}
\end{minipage}
\begin{minipage}[b]{0.5\linewidth}
\subfigure[]{\includegraphics[width=0.43\textwidth]{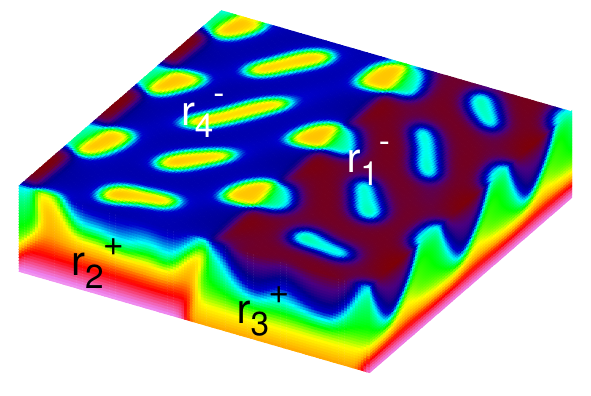} \label{maxEdomain} } \qquad
\subfigure[]{\includegraphics[width=0.43\textwidth]{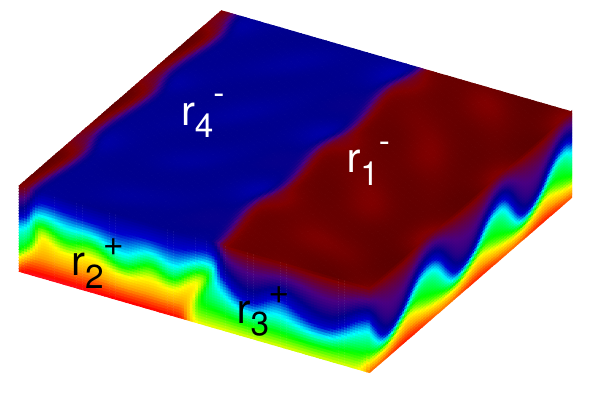}} \\
\subfigure[]{\includegraphics[width=0.43\textwidth]{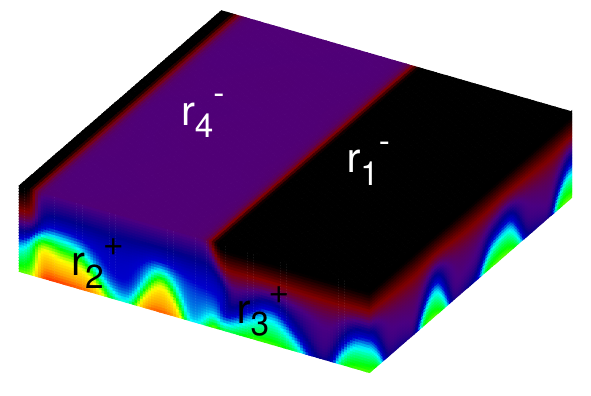} \label{minEdomain} } \qquad
\subfigure[]{\includegraphics[width=0.43\textwidth]{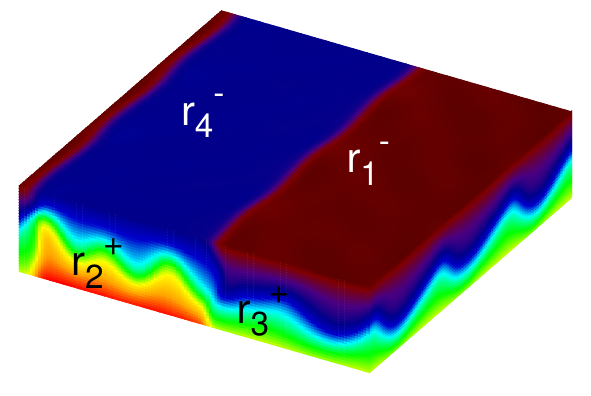}} \\
\end{minipage}
\caption{Simulated hysteresis loop (left) of fully depleted film with $N_D=3 \times 10^{26} \text{m} ^{-3}$, and the domains that form (right two columns) at points (a)-(d) during switching. \label{hys0_05}}
\end{figure*}

Figure \ref{hys0_05} demonstrates these effects for the fully depleted film with a wavy domain wall ($N_D=3 \times 10^{26}$ $\text{m} ^{-3}$) under the same switching conditions that we have applied for the charge-free film in Figure \ref{uncharged}. We also illustrate the domain patterns at maximum magnitude and zero applied fields in the same figure. The magnitude of the average $P_z$ increases with the magnitude of the applied field, and the wavy domain wall moves towards the top or bottom surface of the film. However, not all regions of the domains switch to align with the external field as evident from the presence of darker shaded $r_n^-$ domains remaining at the top surface of the film in Figure \ref{maxEdomain} and lighter shaded $r_n^+$ domains at the bottom surface of the film in Figure \ref{minEdomain}. The internal field induced by the charges at the top of the film opposes the external field under forward bias, and that at the bottom of the film when the field is reverse-biased, leading to incomplete switching of domains. The magnitude of average $P_z$ at maximum magnitude of applied field is thus less than the saturation polarisation of a charge-free film. 

When the external field is removed, the internal field restores the four domains almost to the original state, where the average $P_z$ is close to zero. Therefore the remnant $P_z$ is close to zero and the magnitude of an applied field required to reduce the magnitude of $P_z$ to zero, i.e. the coercive field, is also close to zero; correspondingly, the hysteresis loop narrows. This constriction of the hysteresis loop, as well as the decrease in the remnant polarisation, has been observed in earlier works \cite{Zubko2006, Warren1994}. 

\begin{figure}
\subfigure[]{\includegraphics[width=\columnwidth]{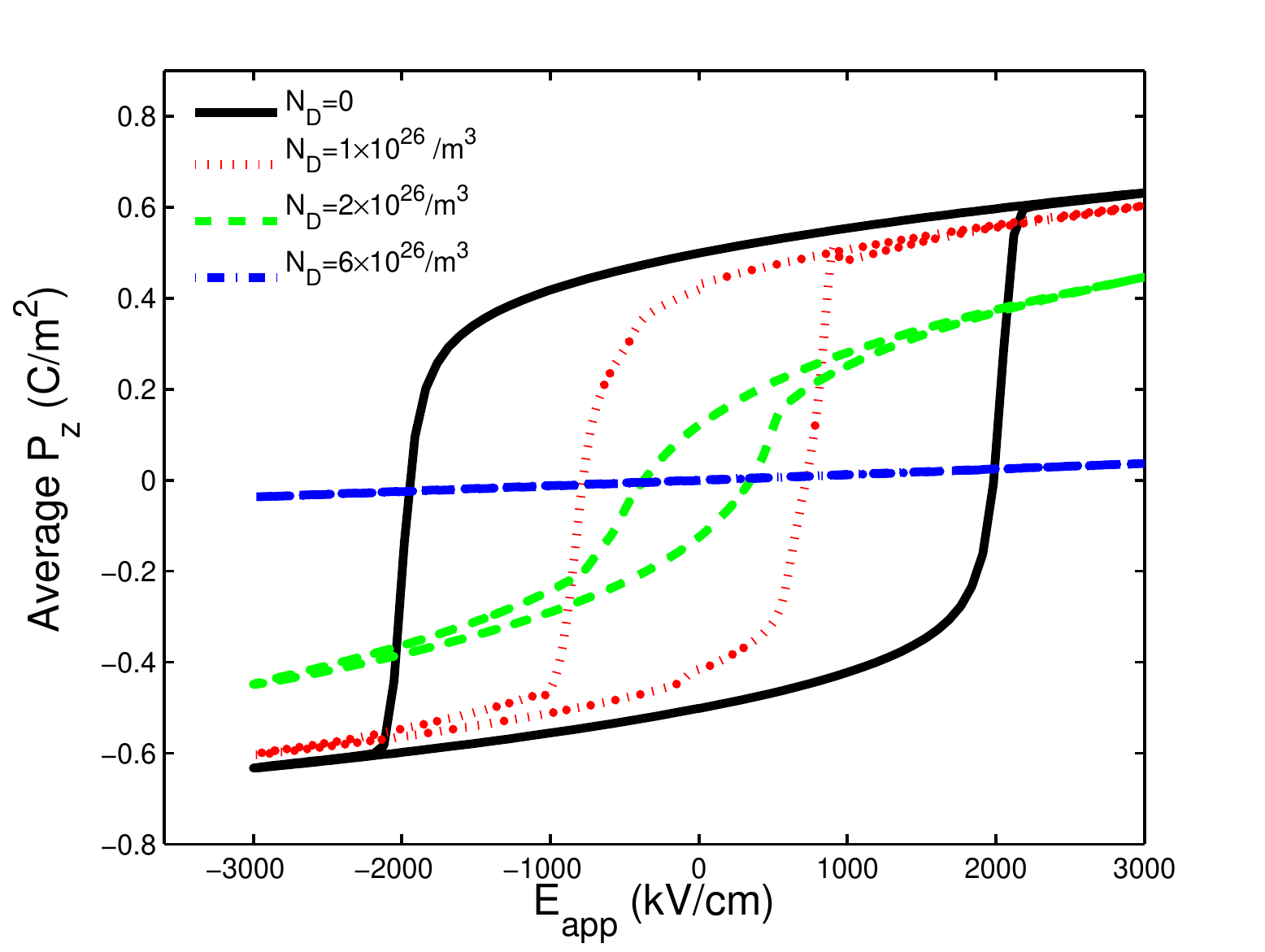} \label{loop_full}} \\
\subfigure[]{\includegraphics[width=\columnwidth]{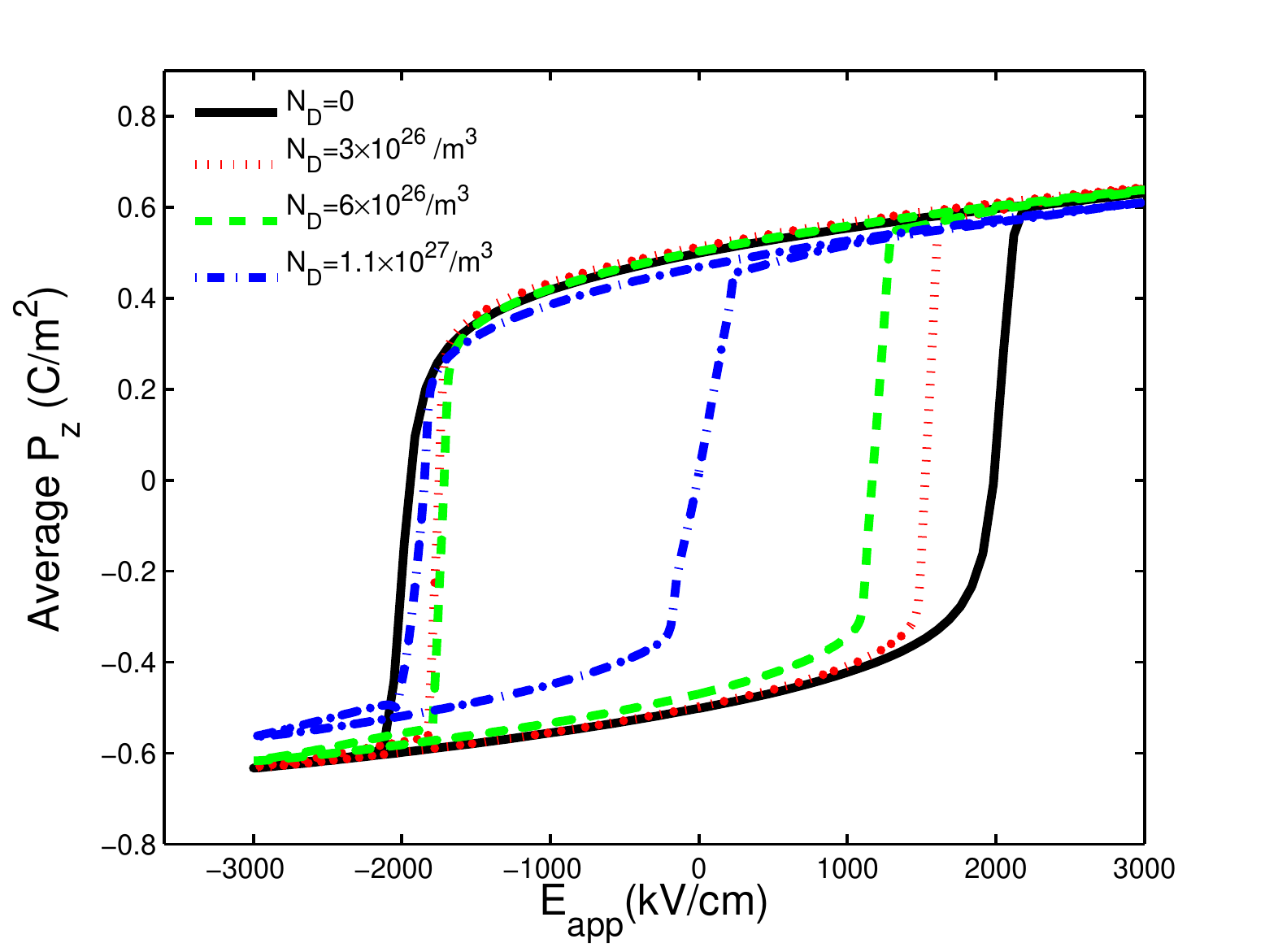} \label{loop_partial}} 
\caption{Simulated hysteresis loops of (a) fully and (b) partially depleted films with $w=5$ nm, for $N_D$ values between $1 \times 10^{26} \text{m} ^{-3}$, to $1.1 \times 10^{27} \text{m} ^{-3}$. The curve for the charge-free film is included for comparison in both graphs. \label{loops}}
\end{figure}

The hysteresis loops for the fully depleted films become more constricted with increasing values of $N_D$, as shown in Figure \ref{loop_full}. This is consistent with the trend observed in other works \cite{Wang2012, Baudry2005}. The increasing magnitude of the internal field in the films increasingly opposes the polarisation switching, such that for films with $N_D\geq 6 \times 10^{26} \text{ m} ^{-3}$ there is no switching of domains but only a slight enhancement in the magnitude of average $P_z$ with $E_{app}$. With very little movement of the domain walls in the $z$ direction, the hysteresis completely disappears. The curve becomes nearly horizontal and linear, reminiscent of that for a paraelectric material.

\begin{figure}
\subfigure[]{\includegraphics[scale=0.32,angle=270,bb=10 300 200 430]{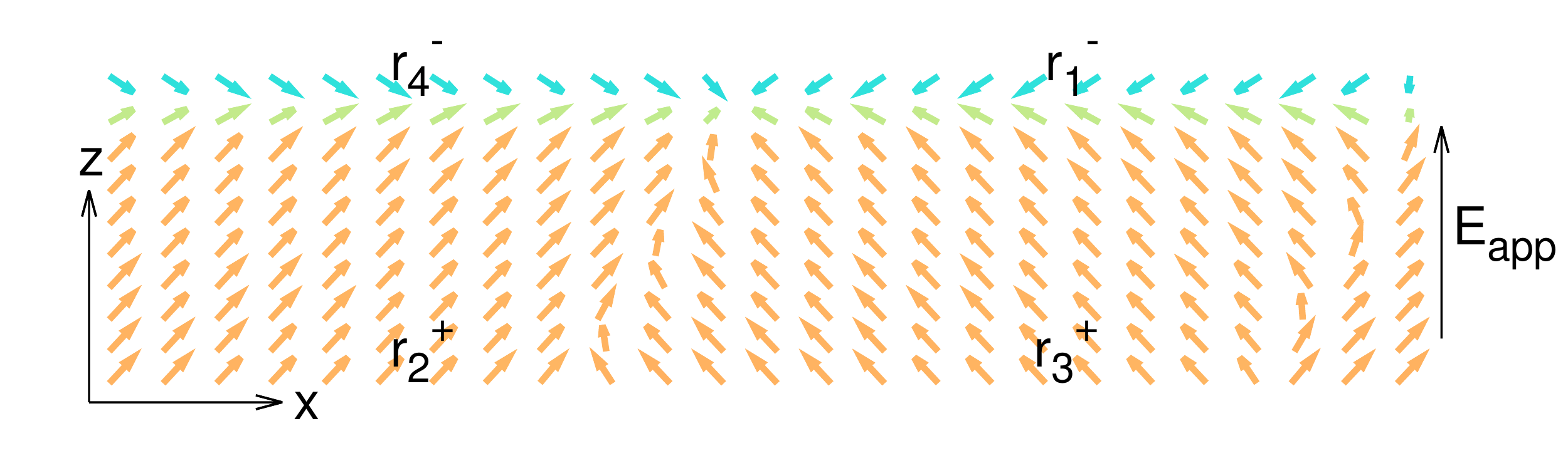}\label{saturation_forward}} \\
\subfigure[]{\includegraphics[scale=0.32,angle=270,bb=10 300 200 430]{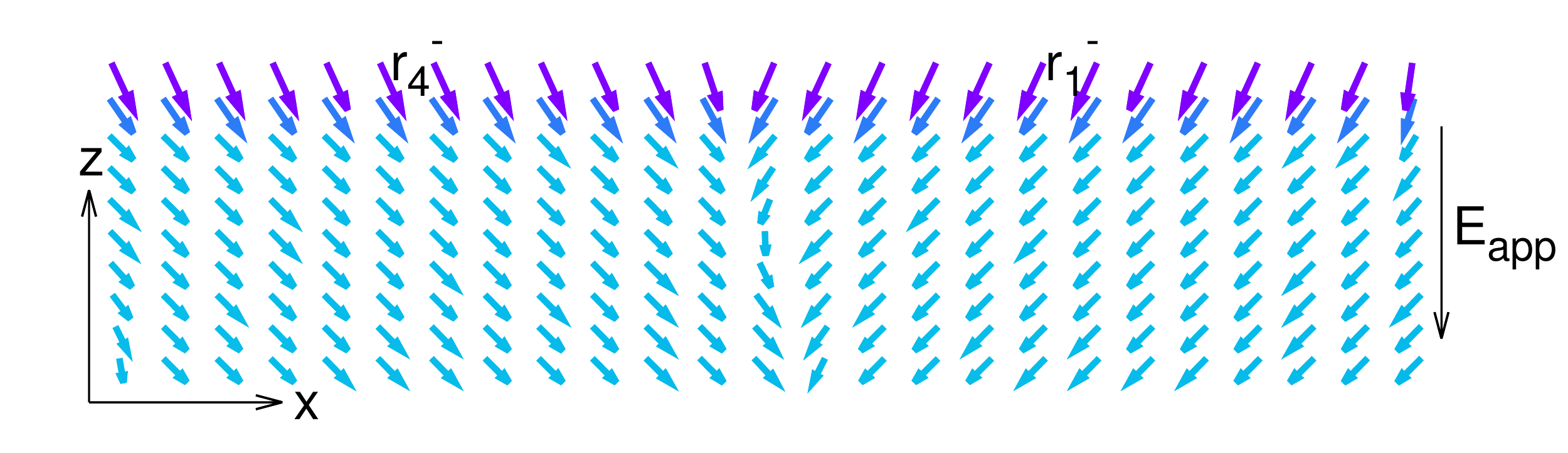}\label{saturation_reverse}}  
\caption{Polarisation vector maps of $P_z-P_y$ in an $x-z$ plane of the partially depleted film with $w=5$ nm and $N_D=1.1 \times 10^{27}$ under an external electric field $E_{app}$ at (a) forward bias and (b) reverse bias. The $r_4^-$ (top left of film) and $r_1^-$ (top right of film) domains indicated in (a) are pinned by the internal field induced within the charged layer under forward bias, whereas the $r_2^+$ (bottom left of film) and $r_3^+$ (bottom right of film) domains switch completely to $r_4^-$ and $r_1^-$ domains respectively with reverse bias. \label{saturation} }
\end{figure}

We also observe a similar trend for partially depleted films, as shown in Figure \ref{loop_partial} for films with $w=5$ nm. In particular, we find that for the partially depleted film with $N_D=1.1 \times 10^{27} \text{ m} ^{-3}$ which forms the four-domain pattern in the absence of external field (Figure \ref{domains}(c)), the $r_1^-/r_4^-$ domains at the top of the charged layer do not switch to $r_3^+/r_2^+$ domains under forward bias (Figure \ref{saturation_forward}), although complete switching of the $r_3^+/r_2^+$ domains to $r_1^-/r_4^-$ domains occurs with reverse bias (Figure \ref{saturation_reverse}). The field induced within the charged layer in this film follows the same trend as those shown in Figure \ref{efield_rho0_1}: the internal field at the top of the charged layer, where $E_z<0$, is much larger in magnitude than elsewhere in the film. It is sufficiently large to impede the switching of the $r_1^-/r_4^-$ domains, whereas the induced field within the $r_3^+/r_2^+$ domains is too small to pin the polarisations. This incomplete switching of domains, which we also observe for other partially depleted films with the head-to-head domain configuration, leads to the observation of non-switchable regions in ferroelectrics \cite{Han2014}.

There is in addition a shift of the hysteresis loop along the $E_{app}$ axis that is not present in that of a fully depleted film, commonly referred to as imprint. Due to the asymmetrical distribution of the charges, the upper region of the charged layer where the internal field $E_z < 0$, is much smaller than elsewhere in the film where $E_z > 0$ (Figure \ref{efield_rho0_1}). In the absence of an external field, the average value of the internal field is positive in the $z$ direction despite the presence of the local negative field of much larger magnitude, and hence domains with $P_z > 0$ will be the preferred polarisation state. Consequently, a larger electric field is required to switch the domains such that an average $P_z < 0$ is produced in the film, compared to that for the reverse. The resulting hysteresis loop is thus left-shifted and becomes further left-shifted with increasing strength of the internal field, as can be observed for the film with $w=5$ nm in Figure \ref{loop_partial}. 

\begin{figure}
\includegraphics[width=\columnwidth,bb=10 10 420 310]{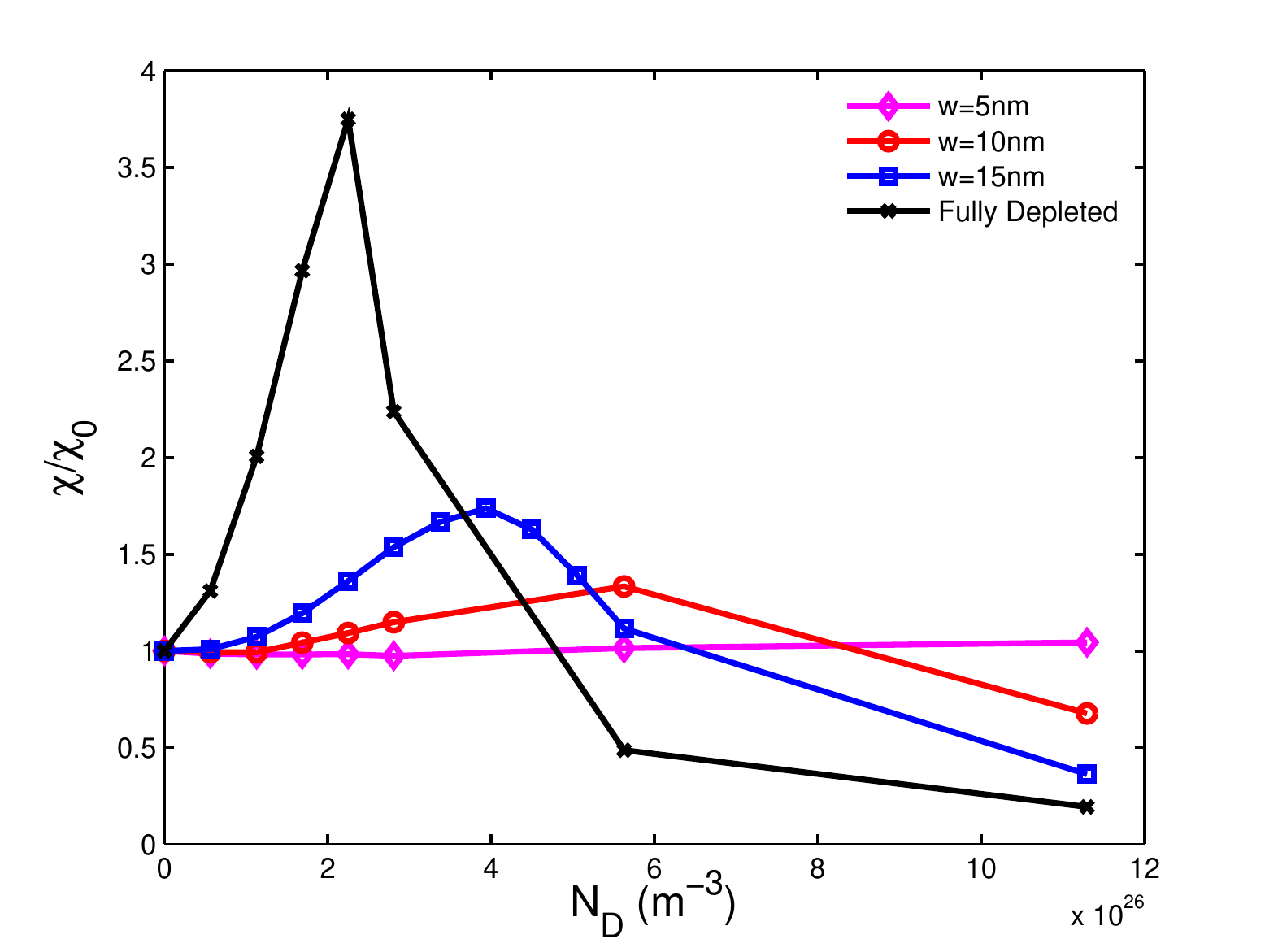}
\caption{$\chi/\chi_0$ of partially and fully depleted films as a function of $N_D$. \label{susceptibility}}
\end{figure}

Having demonstrated the influence of charges on domain patterns and polarisation switching behaviour, we now consider the implications on the dielectric response. From the $P_z-E_{app}$ curve, we calculate $\frac{dP_z}{dE_{app}}$ at zero applied field for $P_z > 0$, which gives an indication of the susceptibility of the film in $z$, $\chi=\frac{1}{\epsilon_0}\frac{dP}{dE}$. Figure \ref{susceptibility} shows this quantity relative to that of the charge-free film, $\chi/\chi_0$, as a function of $N_D$ for partially and fully depleted films. 

\begin{figure*}
\subfigure[$N_D=1.1 \times 10^{26}$ m$^{-3}$]{\includegraphics[width=0.3\textwidth]{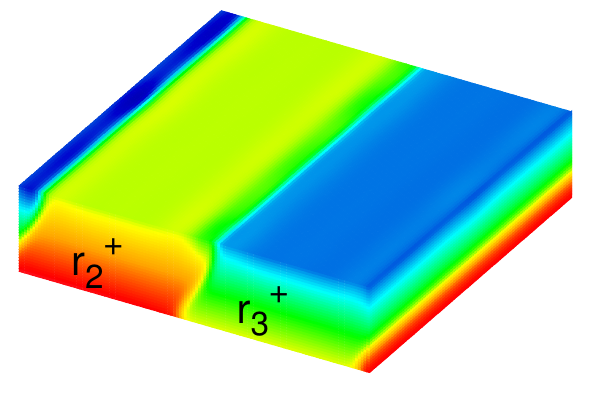}\label{domain_rho0_02}} \quad
\subfigure[$N_D=1.7 \times 10^{26}$ m$^{-3}$]{\includegraphics[width=0.3\textwidth]{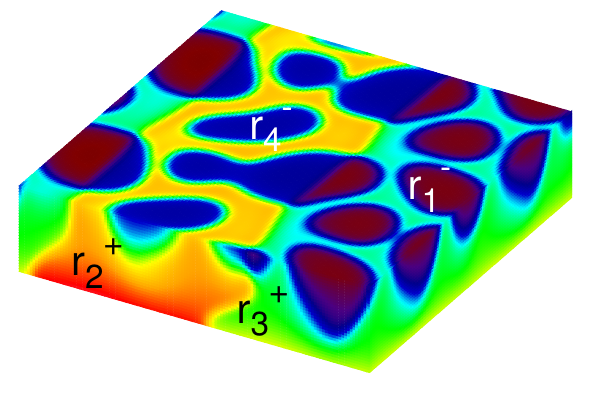}\label{domain_rho0_03}} \quad
\subfigure[$N_D=2.3 \times 10^{26}$ m$^{-3}$]{\includegraphics[width=0.3\textwidth]{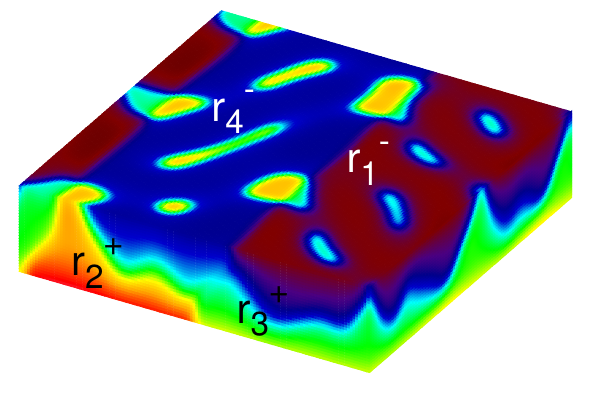}\label{domain_rho0_04}} 
\caption{Domain patterns at remanence ($P_z>0$) of fully depleted films with $N_D$ values from $1.1 \times 10^{26}$ m$^{-3}$ to $2.3 \times 10^{26}$ m$^{-3}$, showing the increasing amount of $r_1^-$/$r_4^-$ domains forming with $N_D$. \label{remanence}}
\end{figure*}

We find that an initial increase in $N_D$ results in increasing enhancement in the susceptibility of the films. 
The magnitude of the internal field at top/bottom surfaces of a charged layer, when large enough, rotates and pins the polarisations from $P_z>0$ to $P_z<0$ in the top region of the charged layer, thus leading to the appearance of head-to-head domains that are stable even at remanence; in comparison, a charge-free film at remanence would have a single polarisation direction in $z$. 
Figure \ref{remanence} illustrates the increasing volume fraction of $r_1^-$/$r_4^-$ domains at remanence with $N_D$. 
As 180\textdegree\ switching of the local polarisations $P_z$ in films with head-to-head domains produces a larger polarisation response compared to that of domains with a single polarisation direction in $z$, the susceptibility of the film therefore enhances with $N_D$.
Additionally we observe that films with higher values of $w$, and thus having larger volume fractions of $r_1^-$/$r_4^-$ domains, show greater enhancement in susceptibility.
The response of the fully depleted film exhibits an increment of almost fourfold when $N_D=2.3 \times 10^{26}$ m$ ^{-3}$, relative to that of the charge-free film. The partially depleted films with $w=15$ nm also show significantly enhanced response, though lower than that of the fully depleted film, of almost twofold when $N_D=4 \times 10^{26}$ m$ ^{-3}$. 
These films have in common wavy interfaces between domains in head-to-head configuration, with hysteresis loops that narrow without complete loss of hysteresis and without significant shifts along the $E_{app}$ axis. 

Below these $N_D$ values, the internal field that acts to pin the domains in the head-to-head configuration is relatively weak in the vicinity of the domain wall, allowing switching of the polarisation with externally applied fields.
However, there is a competing effect also as a result of increasing internal field strength with $N_D$ that inhibits the polarisation switching. The movement of the domain walls becomes more limited and the susceptibility then decreases with further increase in $N_D$, eventually falling to a value that is lower than that of the charge-free film, as can be observed in Figure \ref{susceptibility}.

Furthermore, the maximum in the susceptibility occurs at lower $N_D$ values with increasing $w$, as the magnitude of the internal field that inhibits domain switching also increases with $w$.
For films with suppressed susceptibilities, we have observed that the interface becomes planar and the hysteresis loop becomes almost linear or further left-shifted. 

Thus there exists a range of values for $N_D$ for a given value of $w$ that enhances the susceptibility of a film containing a depletion region, beyond which is detrimental to the susceptibility. This indicates that the presence of space charge in ferroelectric materials may be useful for applications that require high susceptibilities as well as small hystereses. The enhancement due to the presence of charges is in addition to the already known effect of poling along a non-polar direction in a ferroelectric material.

\section{Conclusions}

We have investigated the influence of space charge on domain patterns in rhombohedral ferroelectric thin films, taking \bfo\ as an example. Phase field simulations of domain patterns and polarisation switching have been performed for different charge densities and charged layer widths. We show that the introduction of a charged layer creates an internal electric field which may lead to the formation of double twinned domains in a head-to-head configuration. Depending on the magnitude of the internal field which we vary with charge density and layer width, we find a wavy or planar interface between the head-to-head domains. The former forms in films with intermediate values of $N_D$ above a critical value of $w$, while the latter is observed when $N_D$ is large.

We have also examined the polarisation switching behaviour of the fully depleted as well as the partially depleted films with the applied field along the [001] direction. The widely observed reduction in saturation and remnant polarisation, reduction in coercive field, constriction of the hysteresis loop and, for partially depleted films, its translation along the applied electric field axis, is explained in terms of the internal field that pins or imprints the domain structures as they evolve with the applied switching field. For films in which the charged layer induces a sufficiently large field, the pinning of the domains also results in regions with non-switchable polarisations. We find that the susceptibility of the films can be increased within a range of relatively low $N_D$ values that is high enough to pin domains in a head-to-head (or tail-to-tail) configuration\textemdash this also requires sufficiently thick depletion layer thickness\textemdash but yet low enough to allow movement of domain walls with applied electric fields. In our films, this range of $N_D$ and $w$ coincides with the formation of a wavy interface between head-to-head domains. Our results suggest that space charge-induced domains may engineer small hysteresis, large susceptibility response in ferroelectric materials for device applications.

\begin{acknowledgements}
We are grateful to Khuong P. Ong for useful discussions.
\end{acknowledgements}


\bibliographystyle{ieeetr}
\bibliography{elsarticle-template}

\end{document}